\documentclass[oldversion]{aa}
\usepackage{epsfig}
\usepackage{natbib}
\usepackage{lscape}
\usepackage{wasysym}
\usepackage{amssymb}
\usepackage{epstopdf}
\usepackage{subfigure}

\usepackage{longtable}
\usepackage{rotating}
\usepackage{color}
\usepackage{colortbl}
\usepackage{fancyheadings}
\bibpunct{(}{)}{;}{a}{}{,}

\begin{document}

\title{SOAP-T\thanks{The tool's public interface is available at http://www.astro.up.pt/resources/soap-t/}: A tool to study the light-curve and radial velocity of a system with a transiting planet and a rotating spotted star
}


\author{ M. Oshagh\inst{1,2} \and I. Boisse\inst{1} \and G. Bou{\'e}\inst{1,3} \and M. Montalto\inst{1}
\and N. C. Santos\inst{1,2} \and X. Bonfils \inst{4} \and N. Haghighipour\inst{5}}

\institute{
Centro de Astrof{\'\i}sica, Universidade do Porto, Rua das Estrelas, 4150-762 Porto,
Portugal \\
email: {\tt moshagh@astro.up.pt}
\and
Departamento de F{\'i}sica e Astronomia, Faculdade de Ci{\^e}ncias, Universidade do
Porto,Rua do Campo Alegre, 4169-007 Porto, Portugal
\and
Astronomie et Syst\`emes Dynamiques, IMCCE-CNRS UMR8028,
Observatoire de Paris, UPMC, 77 Av. Denfert-Rochereau,
75014 Paris, France
\and
UJF-Grenoble 1 / CNRS-INSU, Institut de Planétologie et
d'Astrophysique de Grenoble (IPAG) UMR 5274, Grenoble, F-38041, France
\and
Institute for Astronomy and NASA Astrobiology Institute, University of Hawaii-Manoa,
2680 Woodlawn Drive, Honolulu, HI 96822,USA
}

\date{Received XXX; accepted XXX}

\abstract {We present an improved version of SOAP (Boisse et al. 2012) named ``SOAP-T",
which can generate the radial velocity variations and light-curves for systems consisting of a
rotating spotted star with a transiting planet. This tool can be used to study the anomalies inside transit
light-curves and the Rossiter-McLaughlin effect, to better constrain the orbital configuration and properties
of planetary systems and active zones of their host stars. Tests of the code are presented to illustrate its
performance and to validate its capability when compared with analytical models and real data. Finally, we apply
SOAP-T to the active star, HAT-P-11, observed by the NASA Kepler space telescope and use this system to discuss
the capability of this tool in analyzing light-curves for the cases where the transiting planet overlaps with the star's
spots.
}


\authorrunning{M. Oshagh et al.}
\titlerunning{SOAP-T.}
\maketitle

\section{Introduction}

Stellar spots and plages, combined with the rotation of the star, can mimic periodic Radial
Velocity (hereafter RV) signals which may lead to false positives in the detection of a planet \citep{Queloz-01, Bonfils-07, Huelamo-08, Boisse-09,Boisse-11}. Furthermore,
the overlap of the planet and the
active zones on the surface of a star can create anomalies inside the transit light curve
that cause difficulties in constraining the depth of transit and as a result,
the radius of the planet \citep{Rabus-09}. These anomalies can also lead to false positives in the detection of
non-transiting planets by the transit timing variation method \citep{Oshagh-12, Sanchis-Ojeda-11a}.

The stellar spots may provide important information about the orbital configuration of the star's
planetary system. One quantity that is particularly interesting is the angle of stellar spin axis
with respect to the orbital plane of the planet. There are two methods to measure this spin-orbit angle:
the Rossiter-McLaughlin (hereafter RM) effect, which is based on the RV variations of the star
during the planet's transit (e.g., \citealp{Ohta-05, Winn-05, Narita-07, Hebrard-08, Triaud-09, Triaud-10, Winn-10, Hirano-11}),
and the study of the anomalies inside the transit light-curve due to an overlap
of the transiting planet with an active zone of its host star. If long-term and continuous
observations of a star in photometry are accessible (e.g., target stars of the Kepler telescope),
we will be able to detect the evolution of these anomalies inside transits and use them to determine
the spin-orbit angle, inclination of the stellar spin
axis, and the configuration of stellar spots on the surface of the star
(longitude, latitude, and size) \citep{Nutzman-11, Desert-11, Sanchis-Ojeda-11b, Sanchis-Ojeda-12}.
Measuring the angle between stellar spin axis and orbital plane, can give us a better understanding of the configuration of the planetary system as well as
its formation, dynamical evolution, and possible migration
(e.g., \citealp{Queloz-00, Ohta-05, Winn-05, Fabrycky-09, Triaud-10, Morton-11, Hirano-12}).

In this paper, we present a new tool for studying systems consisting
of a rotating star with active zones and a transiting planet. The code, named ``SOAP-T''
generates synthetic radial velocity variations and produces light-curves for the system as
functions of the stellar rotation phase. This tool is fast and as such can be
utilized to simulate different possibilities for different initial configuration in order to
find the best solution for inverse problems.

We describe SOAP-T in section 2. We will explain its software platform, language, input parameters,
methodology, implementation, and output quantities. In section 3 we present
several tests of this software to demonstrate its capabilities. Section 4 has to do with
applying SOAP-T to the particular case of HAT-P-11 and comparing its result with the result
of \citet{Sanchis-Ojeda-11b}. We use this comparison as a demonstration of the capability of SOAP-T in
analyzing the case where the planet and spot are overlapping.  In the last section, we discuss other possible applications of SOAP-T and
conclude this study by presenting its future improvements.

\section{Software description}

SOAP-T has been developed based on the already published code SOAP \citep{Boisse-12}
[http://www.astro.up.pt/soap/] that simulates spots and plages on the surface of a rotating star.
We refer the reader to \citet{Boisse-12} for more details.
The code in SOAP-T is written in Python with calls to functions written in C in order to
make the code faster.

\subsection{Inputs}
The code requires initial conditions which can be assigned by the
user in an input configuration file ``driver.cfg".

The details of the initial parameters for the rotating spotted star can be found in \citet{Boisse-12}.
Here we explain the extra parameters that are required by SOAP-T.\\

\noindent
{\bf Stellar parameters}: One new feature of SOAP-T compared to SOAP is the implementation
of the quadratic limb darkening \citep{Mandel-02} (SOAP considered the linear limb darkening for the star)
\begin{equation}
 I(r)=1-\gamma_{1}(1-\mu)-\gamma_{2}(1-\mu)^{2},
\qquad
\gamma_{1} + \gamma_{2} < 1.
\label{eq:Lebseque I}
\end{equation}
In this equation, $r$ is the normalized radial coordinate on the disk of the star, $I(r)$ is the
specific intensity defined to reach its maximum 1 at the center of the star $(r=0)$
and its minimum at the star's limb, and $\mu= \cos \theta = (1-r^{2})^{1/2}$ where $\theta$
is the angle between the normal to the surface and the observer. As shown by equation (1), the input
parameters require two coefficients of quadratic limb darkening ($\gamma_{1}$ and $\gamma_{2}$).\\

\noindent
{\bf Spot parameters}:
The number of spots is another important parameter. Unlike in SOAP where the number of spots
could be up to 4, it is possible to choose up to 10 spots in SOAP-T.\\

\noindent
{\bf Planet parameters}:

Users are able to set the initial orbital configuration of the planet by
selecting the following quantities: initial value of the planet's orbital period $(P_{orb})$,
the time of its periastron passage $(T_{0})$, its initial orbital phase $(PS_{0})$,
eccentricity $(e)$, semimajor axis $(a)$, argument of periastron  $(\omega)$, inclination
with respect to the line of sight $(i)$, as well as the planet's radius $(R_{planet})$ and the angle
between the stellar spin axis and the normal to the orbit of the planet projected on the $y-z$ plane. We denote this
angle by $\lambda$.

\subsection{Detail description of computation}

We restate here from \citet{Boisse-12} that the rotating spotted star is centered on a
grid of $grid \times grid$ cells on a y-z plane. The cells take their y and z values
between -1 and 1, normalized to the stellar radius. Each grid cell ($p_{y}, p_{z}$)
contains a flux value and a CCF (cross correlation function). Our
goal in this section is to describe 1) how we add a planet to the rotating spotted star,
2) how we compute its impact in photometry and in RV, and 3) how we incorporate the overlap
between the stellar inhomogeneities and the planet.

To determine the effect of a planet on the RV and flux of a non-spotted star in each phase step,
the code first checks whether the planet is
in the foreground or the background with respect to the star. For that reason,
in the first step, SOAP-T calculates the position of the planet for each time step.
Since the output of the code is in the stellar rotational phase, the code uses the following
equation to convert stellar rotation phase into time
\begin{equation}
 T = (PS + PS_{0})\times P_{rot}.
\label{eq:Lebseque I}
\end{equation}
The quantity $PS$ is the stellar rotation phase which can be between 0 and 1,
$PS_{0}$ is the planet's orbital initial phase which defines the initial
position of the planet, and $P_{\rm rot}$ is the stellar rotation period in days.
This equation is only valid for stellar rotation periods larger than the orbital period of the planet,
which should be true for the most cases of transiting extrasolar planetary systems.

In order to determine the planet's position, the code has to calculate $\lambda$.
As explained above, $\lambda$ is the angle between the stellar spin axis (denoted by ${\hat n}_\ast$) projected on the $y-z$ plane and the unit vector normal to the plane of the planet's orbit
(shown by ${\hat n}_p$ in figure 1) also projected on
the same plane. The coordinates of vector ${\hat n}_\ast$ are determined using

\begin{eqnarray}
 {{\hat n}_{\ast}}= R_{x}(\psi_{\ast}) R_{y}(i_{\ast})\left(\begin{array}{c}\nonumber
 1\\\nonumber
0\\\nonumber
0\end{array} \right)\nonumber = \left(\begin{array}{c}\nonumber
 \cos i_{\ast}\\\nonumber
\sin i_{\ast}\sin \psi_{\ast}\\\nonumber
-\sin i_{\ast}\cos \psi_{\ast}\end{array} \right),\\
\end{eqnarray}

\noindent
where $(\psi_{\ast})$  is the longitude of stellar spin, $(i_{\ast})$ is the star's inclination, and
$R_x$ and $R_y$ are the rotation matrices around the $x$ and $y$ axes.
The coordinates of the normal vector ${\hat n}_p$ are also determined in the same fashion
using the planet's longitude of ascending node  $(\Omega)$ and orbital inclination $(i)$ as

\begin{eqnarray}
 {{\hat n}_{p}}= R_{x}(\Omega) R_{y}(i)\left(\begin{array}{c}\nonumber
 1\\\nonumber
0\\\nonumber
0\end{array} \right)\nonumber = \left(\begin{array}{c}\nonumber
 \cos i\\\nonumber
\sin i\sin \Omega\\\nonumber
-\sin i\cos \Omega\end{array} \right).\\
\label{eq:Lebseque I}
\end{eqnarray}

\noindent
The angle $\lambda$ is then calculated using $\lambda=\Omega - \psi_{\ast} $. The code, subsequently,
aligns the projected stellar spin axis with the $z$ direction by setting $\psi_{\ast}={180^\circ}$ and uses
the longitude of the ascending node of the planet $\Omega= 180+ \lambda$ to calculate the coordinates of
its position vector (see section 4 and equations 53-55 in Murray \& Correia 2011 for more details).
At each stellar phase step, the position vector of the planet ($\textbf{D}$) is given by

\begin{eqnarray}
\textbf{D}=r_{p}\left(\begin{array}{c}\nonumber
\sin i \sin(\omega+f)\\\nonumber
-\cos \lambda \cos (\omega+f)+ \sin \lambda \cos i \sin (\omega+f)\\\nonumber
-\sin \lambda \cos (\omega+f)- \cos \lambda \cos i \sin (\omega+f) \end{array} \right).\\
\end{eqnarray}

\noindent
In equation (5)

 \begin{equation}
r_{p}=\frac{a(1-e^{2})}{1+e \cos f},
\label{eq:Lebseque I}
\end{equation}
\noindent
where $r_{p}$ is the distance between the stellar center and the center of planet, and

\begin{equation}
f = 2\tan^{-1} \left[{\left({\frac{1+e}{1-e}}\right)^{1/2}} \tan \left(\frac{E}{2}\right)\right]
\label{eq:Lebseque I}
\end{equation}
is the true anomaly of the planet. To calculate $f$, for small eccentricity, the code uses
Kepler's equation
\begin{equation}
E=M+\frac{e \sin M - M}{1-e\cos M},
\label{eq:Lebseque I}
\end{equation}

\noindent
where $M=2\pi(T-T_{0})/P_{orb}$ is the mean-anomaly of the planet's orbit (e.g. \citet{Ohta-05, Murray-11}) (see Figure 1, a schematic view of a planetary system).

\begin{figure*}[t]
\center
\includegraphics[scale=0.7]{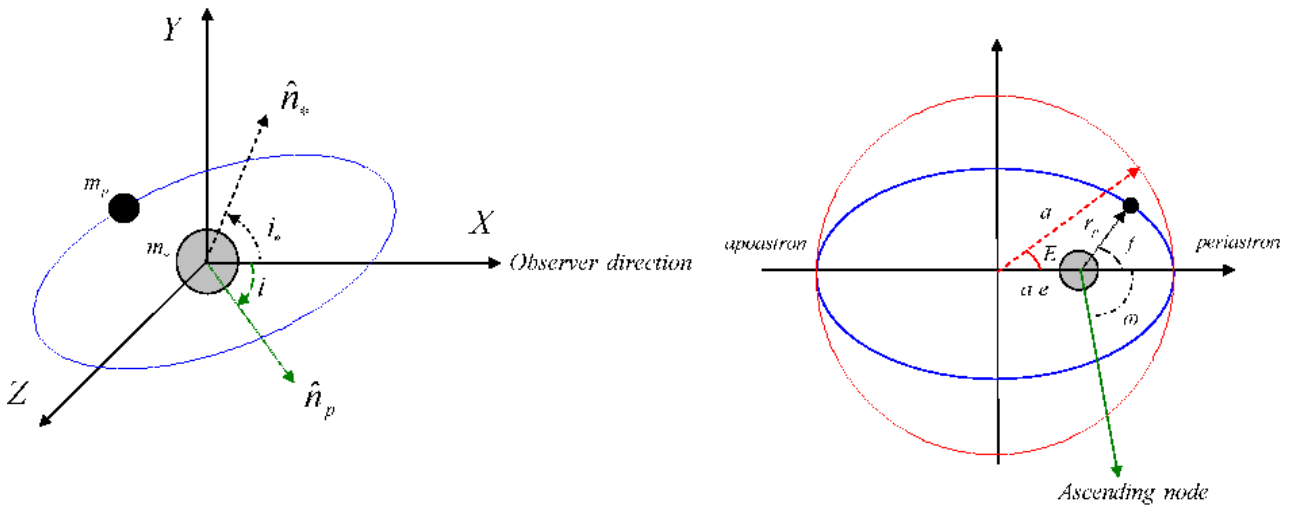}
\caption{Left: Schematic configuration of the stellar spin axis and planetary orbital plane. The vector
$\hat{n}_{p}$ is the normal unit vector of the planetary orbital plane, the vector
$\hat{n}_{\ast}$ is the stellar spin axis, and $i_{\ast}$ is the stellar
inclination. We denote the angle between the projected stellar spin axis and the projected normal unit vector of the
planetary orbital plane on the $y-z$ plane with $\lambda$
(not shown here). Right: Schematic view of planetary orbit from above the orbital plane, the star is the gray circle and the
planet is the black full circle. See Section 2 for more details on the symbols.}

\end{figure*}

If the planet is in the foreground ($\textbf{D}_{x} > 0$), the code checks if the planet is inside
the stellar disk or outside by comparing the projected distance of the planet's center to the
stellar center with the radius of star ($|\textbf{D}|< R_{\ast}$). If the planet is inside,
the code resolves the area of the grid where the planet is located and establishes an area around the planet where it will focus the subsequent
calculations. This area is then scanned to determine whether each grid cell is located inside
or outside the stellar disk (especially in the case where the planet is on the stellar limb), and if
this grid cell belongs to the planet's disk, the code determines the location by calculating the distance
of the grid cell to the center of planet and comparing it with planet's radius.
Assuming that a grid cell is located inside the stellar disk and also belongs
to the planet's disk, its CCF is modeled by a Gaussian with a width and amplitude given by the input
parameters and doppler-shifted according to the projected stellar rotation velocity weighted
by the quadratic limb-darkening law. This CCF value will then be removed from the total CCF of
the rotating non-spotted star. For the same grid cell, the values of the flux can be calculated using only
on the quadratic limb-darkening law. The code also removes the flux value of that cell from the total flux of the
rotating spotted star.

In the next step, the code adds a planet to the rotating spotted star in order to calculate the
effect of both the planet and the spots on the RV and flux at the same time. In the case that the
spot is not partly covered by the planet, the procedure is as explained below. The code
\begin{itemize}
\item determines the position of the spot,
\item determines the position of the planet,
\item applies inverse rotation to the spot to bring it along the line of sight,
\item scans the region of the spot's position to identify the grid cells that belong to the spot's disk area,
\item sums up the CCF and flux of all cells inside the spot and removes them
from the total flux and CCF of the non spotted-star [see figure 4 of Boisse et al. (2012) for the inverse rotation and also
for more
details on spots calculation],
\item scans the region of the planet's position to determine the grid cells that are inside the
disk of the planet,
\item calculates the sum of the CCF and flux for those grid cells inside the transiting planet and
removes them from the total flux and CCF of the spotted star.
\end{itemize}

\noindent
The complexity arises from the overlapping of the transiting planet
with a spot on the surface of the star. In this case, the code considers those parts of the spot
whose distances to the center of the planet are smaller than the planet's radius as the overlapped area. Those points will
not be scanned during the spot scanning process. As a consequence, they will skip the rest
of procedure for CCF and flux of spot and removal from total CCF and flux. This procedure is able to produce those positive
``bump" anomalies inside the transit light curve that are due to the reduction in the loss of light.
The schematic view of the system which is simulated by SOAP-T is shown in Figure 2.

\subsection{Outputs}

The code returns the results of its simulation into the output file ``output.dat".
The output file contains four columns that represent stellar phase, flux, RV and BIS (the bisector span), respectively.
These quantities can be plotted
directly by SOAP-T as functions of the stellar rotation phase. Similar to SOAP, the stellar phase step can be
chosen in two ways, first by selecting the constant fraction of the phase between 0 to 1, or by
uploading the wanted phase values in the $ph\_in$ file. The flux values correspond to
the photometry of the system (the total flux of system), and they are normalized to the maximum value of total flux during the stellar rotation phase. The RV values are obtained by fitting a Gaussian
function to each total CCF [see more details in \citet{Boisse-12}]. We note here that each total
CCF is also normalized to its maximal value. We refer the reader to \citet{Queloz-01} for details on BIS
calculations.

\section{Tests}

To check the capability of SOAP-T and the validity of its result, we preformed several tests to
compare its outcome with theoretical models and real observations. Tests related
to the effects of spots on the RV and flux signals have been presented and explained in
\citet{Boisse-11, Boisse-12}. In this section, we test the validity of the results for the photometry
and RV with a transiting planet around both a non-spotted and a spotted star.

To begin with, we consider a system consisting of a non-spotted star with a transiting planet.
The purpose of this test is to determine whether the code can reproduce the light curve of a star with a
transiting planet as obtained from the theoretical model of \citet{Mandel-02}.  We select
various initial conditions including a star without any limb darkening,
or with linear or quadratic limb darkening. As shown in Figure 3, the
results obtained from SOAP-T are in strong agreement with the results obtained from the theoretical
model based on the formalism presented by \citet{Mandel-02}. The latter implies that SOAP-T can be used
reliably to characterize the light-curve of a transiting planetary system.

In a second test, we examined the RVs obtained using SOAP-T during the transit of a planet in front of
a non-spotted star. We applied SOAP-T to the system of WASP-3 which
is known to harbor a planet with a misalignment between its orbital plane and stellar spin axis
(see the parameters of WASP-3 system in the Table 1). The RV observation data of  WASP-3 during its
transits were taken from \citet{Simpson-10} (it was observed with the SOPHIE spectrograph at Haute-Provence Observatory). We fitted the RVs produced by SOAP-T to the observational data and compared the results with those of \citet{Simpson-10}. We only allowed stellar rotation velocity
$(v \sin i)$ and misalignment angle ($\lambda$) to vary as free parameters, and kept all other parameters
constant and equal to their values reported by \citet{Gibson-08} and \citet{Simpson-10} (see
Table 1).

The best fit result obtained using SOAP-T corresponds to
$\lambda= 20.0^{\circ} \pm 3.3^{\circ}$ and $v\sin i= 13.31 \pm 0.45 \, {\rm km s^{-1}}$
with $\chi^{2}_{\rm red}=0.9308$. We obtained the error bars of these measurements using the bootstrap
method \citep{Wall-03}. The value of the misalignment angle $(\lambda)$ is consistent with the value reported by
\citet{Simpson-10} ($13^{\circ}\pm^{9^{\circ}}_{7^{\circ}}$) within $1\sigma$. Note that the error bars obtained with SOAP-T are more than twice smaller than those derived by \citet{Simpson-10}. We assume that this could be due to smaller number of free parameters in our fitting procedure, otherwise it might be also due to the fact that SOAP-T has been developed to reproduce exactly the analysis routine of RV measurements and is thus closer to the observations when compared to the result of the analytical formula used by \citet{Simpson-10} \citep{Boue-12}. Indeed, we obtained also a better agreement between the projected stellar rotation velocity $(v\sin i)$ and the value obtained by spectroscopic broadening measurements ($13.4\pm 1.5 kms^{-1}$). For comparison, \citet{Simpson-10} obtained $v\sin i=19.6 \pm ^{2.2}_{2.1} kms^{-1}$. Figure 4 shows the result of the best SOAP-T RV's fit and the observation of RM effect of WASP-3.

\begin{table*}[htdp]\scriptsize

\caption{The stellar and planet's parameters of the WASP-3 system \citep{Gibson-08, Simpson-10}.}
\begin{center}
\begin{tabular}{c c c c c c c}

\hline
Parameter & Value  & Uncertainty  \\

\hline
\textbf{Stellar parameters}\\
$R_{\star} (R_{\bigodot})$ & 1.31 & $\pm ^{0.05}_{0.07}$ \\
Linear limb darkening coefficient & 0.69 & - \\
Stellar inclination (deg) & 90 & - \\
Spectroscopic stellar rotation velocity $v\sin i (kms^{-1})$ & 13.4 & $\pm 1.5$ \\

\hline
\textbf{Planet parameters}\\
Planet to star radius ratio $(R_{p}/R_{\ast})$ & 0.1013 & $\pm ^{0.0014}_{0.0013}$ \\
Period (days)& 1.846835 & $\pm 0.000002$ \\
Eccentricity & 0 & - \\
Scaled Semimajor axis $(a/R_{\ast})$ & 5.173 &  $\pm ^{0.246}_{0.162}$ \\
Orbital inclination (deg) & 84.93 & $\pm ^{1.32}_{0.78}$ \\
Projected spin-orbit misalignment angle(deg) &  13 & $\pm ^{9}_{7}$ \\

\hline
\end{tabular}
\end{center}
\label{default}
\end{table*}%

Finally, since there is no theoretical model for the case where a transiting planet and star spots
overlap, to evaluate the performance of SOAP-T in this case, we compared the photometric result
of SOAP-T with the results reported in the literature
for the real observation of HAT-P-11 by the Kepler space telescope. We will explain this test and
its applications in the next section.

\begin{figure}[t]
\center
\includegraphics[trim=0mm 0mm 0mm 0mm, clip,width= 10 cm]{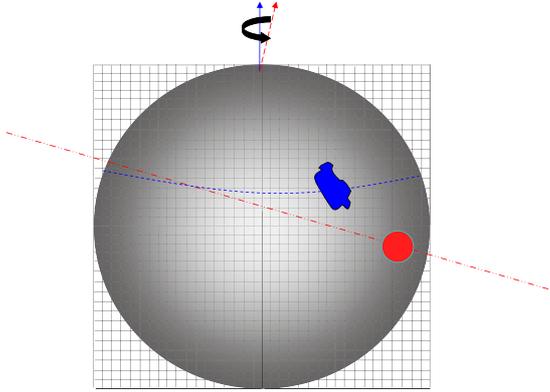}
\caption{Schematic view of stellar spot and the transiting planet in simulations by SOAP-T.}
\label{fig:exslope}
\end{figure}

\begin{figure}[t]
\center
\includegraphics[scale=0.45]{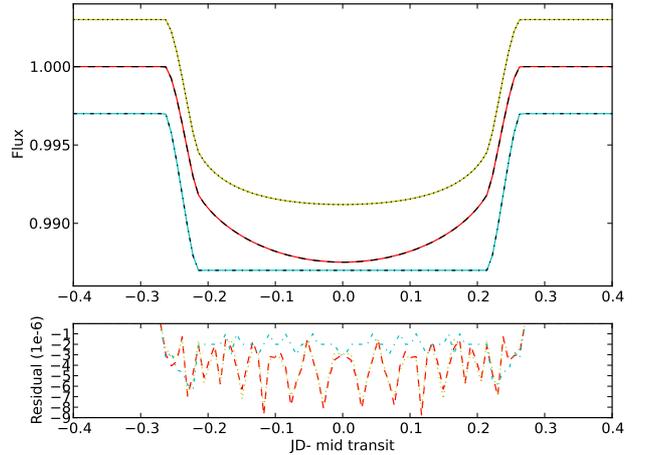}
\caption{Comparing the photometric results of SOAP-T and the theoretical model of a
transiting planet over a non-spotted star \citep{Mandel-02}. The cyan, red, and yellow lines show SOAP-T's results for a
star without limb darkening, a star with linear limb darkening $({\gamma_1}= 0.6)$, and a star with quadratic
limb darkening ($\gamma1=0.29$ and $\gamma2=0.34$), respectively. The dash-dotted line, dashed line, and the dotted line show
the corresponding results using the mechanism by \citet{Mandel-02}.}
\label{fig:exslope}
\end{figure}

\begin{figure}[t]
\center
\includegraphics[scale=0.5]{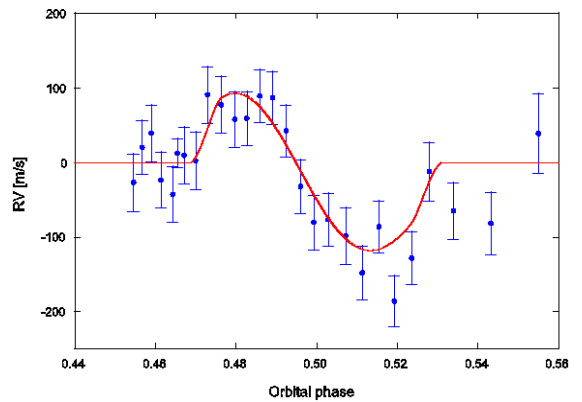}
\caption{RV observations of WASP-3 during transit of WASP-3b (RM effect) minus Keplerian motion, which are blue dots \citep{Simpson-10}, overplotted with the best fit of RV signals obtained from SOAP-T, which is red line.}
\label{fig:exslope}
\end{figure}

\begin{figure}[t]
\center
\includegraphics[scale=0.45]{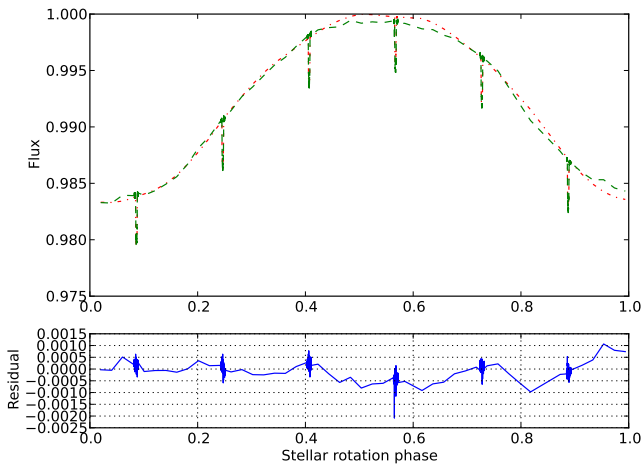}
\caption{Comparing the best fit model to the transit photometry of HAT-P-11 using SOAP-T and the Kepler observation of HAT-P-11 for
one period of stellar rotation. Red (dash-dotted) line shows SOAP-T's photometric result for an edge-on solution
and green(dashed) line correspond to HAT-P-11 observation. The blue (solid) line in the bottom
panel shows the residual.}
\label{fig:exslope}
\end{figure}

\section{Application of SOAP-T to HAT-P-11}

HAT-P-11 is a 6.5 $Gyr$  bright (V=9.6) K4 star with a mass of 0.81 $M_{\bigodot}$ and radius
of 0.75 $R_{\bigodot}$. At a distance of 38 pc, this star hosts a transiting Neptune-sized
planet with a  mass of 0.081 $M_{Jup}$ and  radius
of 0.422 $R_{Jup}$ \citep{Bakos-10}. HAT-P-11b has a period of 4.887 days and
revolves around its central star in an orbit with a semimajor axis of 0.053 AU and an
eccentricity of $e= 0.198$. More details of the parameters of this planet and its host
star can be found in Table 2.

Studies of the HAT-P-11 planetary system have shown extreme misalignment of the projected angle between the stellar spin axis and the planet's orbital
plane $(\lambda= {103^\circ \pm^{26^{\circ}}_{10^{\circ}}})$ through RM effect \citep{Winn-10}.

HAT-P-11 (KOI-3) was used as one of the calibration target stars for the Kepler space telescope
and has been observed throughout Kepler's mission in both short and long cadences. The high precision
photometry of this star has revealed some features inside its transit light curve
and has also shown a large modulation outside of transits. The anomalies
inside the transit light curve of HAT-P-11 can be attributed to the overlapping of the transiting
planet with active zones on the surface of the star \citep{Sanchis-Ojeda-11b}.
The recurrence (or not) of these anomalies can be used to put an upper limit on the stellar obliquity.
Recently \citet{Sanchis-Ojeda-11b} carried out
a purely photometric study of this system based on the observation of photometric anomalies inside
the transit light curve using the Kepler public data Q0-Q2 and showed that
no recurrence of anomalies exists in two closely spaced transits in
the Kepler data of HAT-P-11. They considered this result as evidence of
misalignment of the stellar spin axis with respect to orbital plane. Fortunately,
these authors were able to detect a feature
with two peaks in the folded light curve of 26 transits of this star which could be interpreted as evidence for the existence of two long lived spot belt
regions on the star. By using a simple geometric model, these authors arrived at two solutions for the
stellar obliquity; an edge-on solution that is in good agreement with the result of \citet{Winn-10} through
the RM study, and an alternative solution in which the star is seen almost pole-on.
The method used by \citet{Sanchis-Ojeda-11b} also allowed these authors to put some constraints on the
position (latitude) and size of the active zones on the surface of star (see details of parameters
for two solutions in Table 3).

\begin{table*}[htdp]\scriptsize

\caption{The stellar and planet's parameters of the HAT-P-11 system \citep{Bakos-10, Sanchis-Ojeda-11b}.}
\begin{center}
\begin{tabular}{c c c c c c c}

\hline
Parameter & Value  & Uncertainty  \\

\hline
\textbf{Star parameters}\\
$M_{\star} (M_{\bigodot})$ & 0.81  & $\pm ^{0.02}_{0.03}$ \\
$R_{\star} (R_{\bigodot})$ & 0.75  & $\pm 0.02$ \\
Stellar rotation period (days) & 30.5 & $\pm ^{4.1}_{3.2}$ \\
Age (Gyr) & 6.5  & $\pm ^{5.9}_{4.1}$ \\
Distance (pc) & 38.0  & $\pm 1.3$ \\
Linear limb darkening coefficient  & 0.599  & $\pm 0.015$ \\
Quadratic limb darkening coefficient  & 0.073  & $\pm 0.016$ \\

\\
\textbf{Planet parameters}\\
$M_{P} (M_{Jup})$ & 0.081  & $\pm 0.009$ \\
Planet to star radius ratio $(R_{p}/R_{\ast})$ & 0.05862 & $\pm0.00026$\\
Orbital period (days) & 4.8878049  & $\pm 0.0000013$ \\
Scaled Semimajor axis $(a/R_{\ast})$ & 15.6  & $\pm 1.5$ \\
Eccentricity & 0.198  & $\pm 0.046$ \\

\hline
\end{tabular}
\end{center}
\label{default}
\end{table*}%

Here we present the results of the application of SOAP-T to HAT-P-11
in order to reproduce the light curve of this star
over one stellar rotation. The rotation period of this star contains six transits.

Since the inside transit anomalies carry more information about the configuration of a planetary system,
we decided to give more weight to these points in our study. We, therefore,
performed a regular sampling to reduce the number of points outside of
the transit light curve. One point was taken from each 300 points in the light curve out side the
transit. This process reduced the running time of the code significantly.

\subsection{HAT-P-11 Edge-on solution}
In this section, we consider the system to be edged-on and
choose all initial parameters of the star, planet, and spots (except the number of spots and their
longitude) from the values reported by
\citet{Sanchis-Ojeda-11b}. The spot's brightness are fixed to zero. The details of initial conditions are listed
in Table 3. To reproduce all the features in the light curve,
both inside and outside the transit,  we limited the latitude and the size of the spots to
the range reported by \citet{Sanchis-Ojeda-11b}. The spots's radius are equal to the half-width of the active zone which correspond to $R_{spot}= 0.08 R_{\ast}$). However, we allowed the number of spots and their
longitudes to vary as free parameters between 0-10 and ${0^\circ}-{360^\circ}$, respectively.
The best fit to the observations ($\chi^{2}_{red} = 7.356 $) was obtained with eight spots on the
surface of the star. Figure 5 shows the best result of SOAP-T photometry overplotted with the real observation of HAT-P-11 during one stellar period. Figure 6 shows more details of the
inside of each transit and the results obtained from SOAP-T after normalizing
both flux values to one. As shown in both figures, SOAP-T could manage to reproduce all the inside transit anomalies and also the out side transit variation. We note here that the results of
\citet{Sanchis-Ojeda-11b} were obtained by considering the inside transit anomalies.
These authors did not consider the outside features of the light curve.
However, using SOAP-T we were able to study the entire light-curve, both inside transit
and outside, at the same time. We note that here we do not intend to confirm the results of
\citet{Sanchis-Ojeda-11b}. We only use the similarities between the results obtained
from SOAP-T photometry (e.g. position and size of spots, stellar inclination and spin-orbit misalignment angle) and the results by these authors as a proof of the validity of SOAP-T's capability in simulating the case of spot and planet overlapping.

\begin{table*}[htdp]\scriptsize

\caption{Parameters of the HAT-P-11 system for different solutions reported by
\citet{Sanchis-Ojeda-11b}.}
\begin{center}
\begin{tabular}{c c c c c c c}

\hline
Parameter & Value  & Uncertainty  \\

\hline
\textbf{Edge-on solution}\\
Projected spin-orbit misalignment angle $\lambda$ (deg) & 106  & +15/-12 \\
Stellar inclination $i_{s}$ (deg) & 80  & +4/-3 \\
Latitude of active zone (deg) & 19.7  & +1.5/-2.2 \\
Half-width of active zone (deg) & 4.8  & +1.5/-1.8 \\

\\
\textbf{Pole-on solution}\\
Projected spin-orbit misalignment angle $\lambda$ (deg) & 121  & +24/-21 \\
Stellar inclination $i_{s}$ (deg) & 168  & +2/-5 \\
Latitude of active zone (deg) & 67  & +2/-4 \\
Half-width of active zone (deg) & 4.5  & +1.6/-1.9 \\

\hline
\end{tabular}
\end{center}
\label{default}
\end{table*}%

\begin{figure*}[hbtp]
  \vspace{9pt}

  \centerline{\hbox{ \hspace{0.0in}
    \epsfxsize=2in
    \epsffile{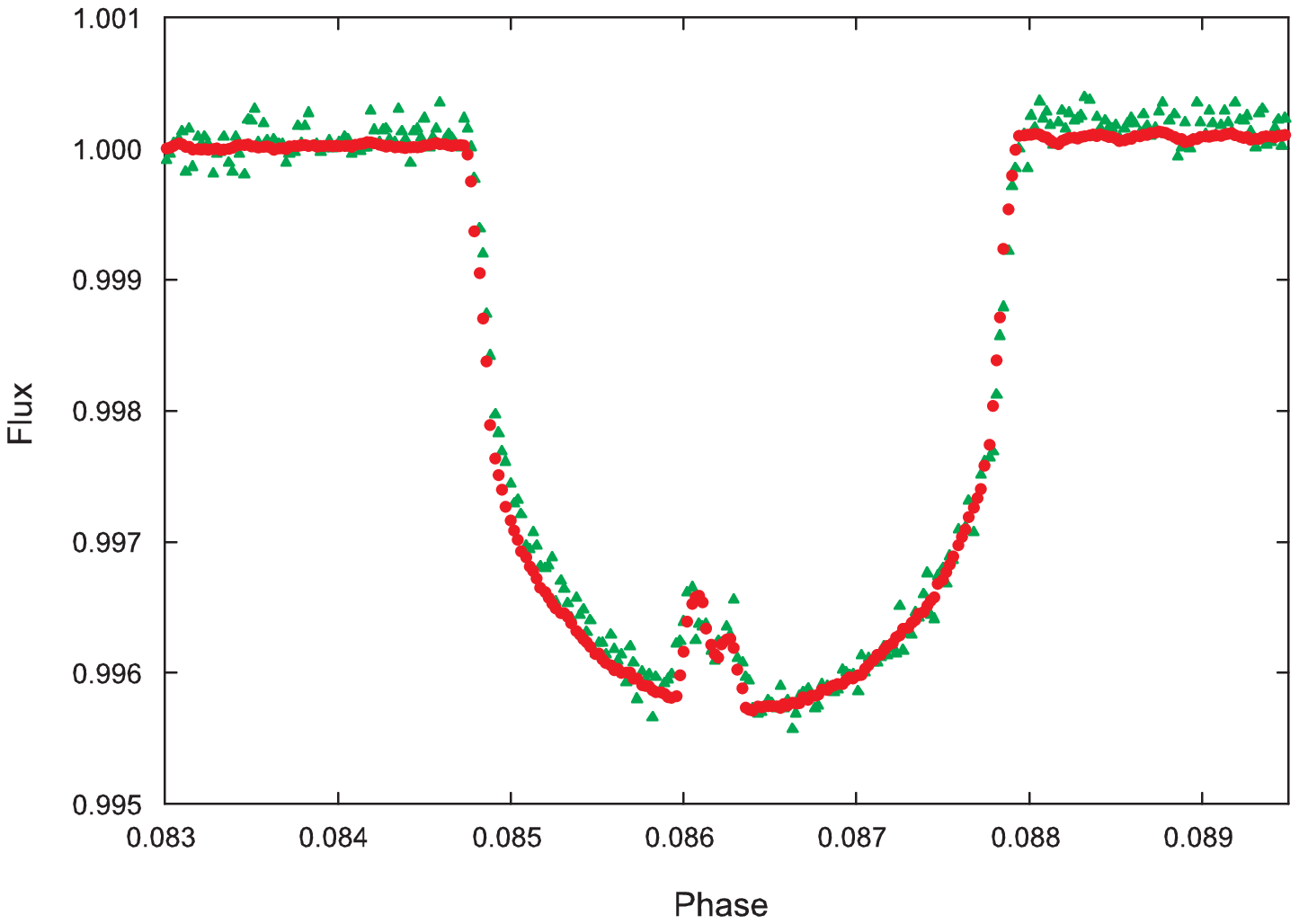}
    \hspace{0.05in}
    \epsfxsize=2in
    \epsffile{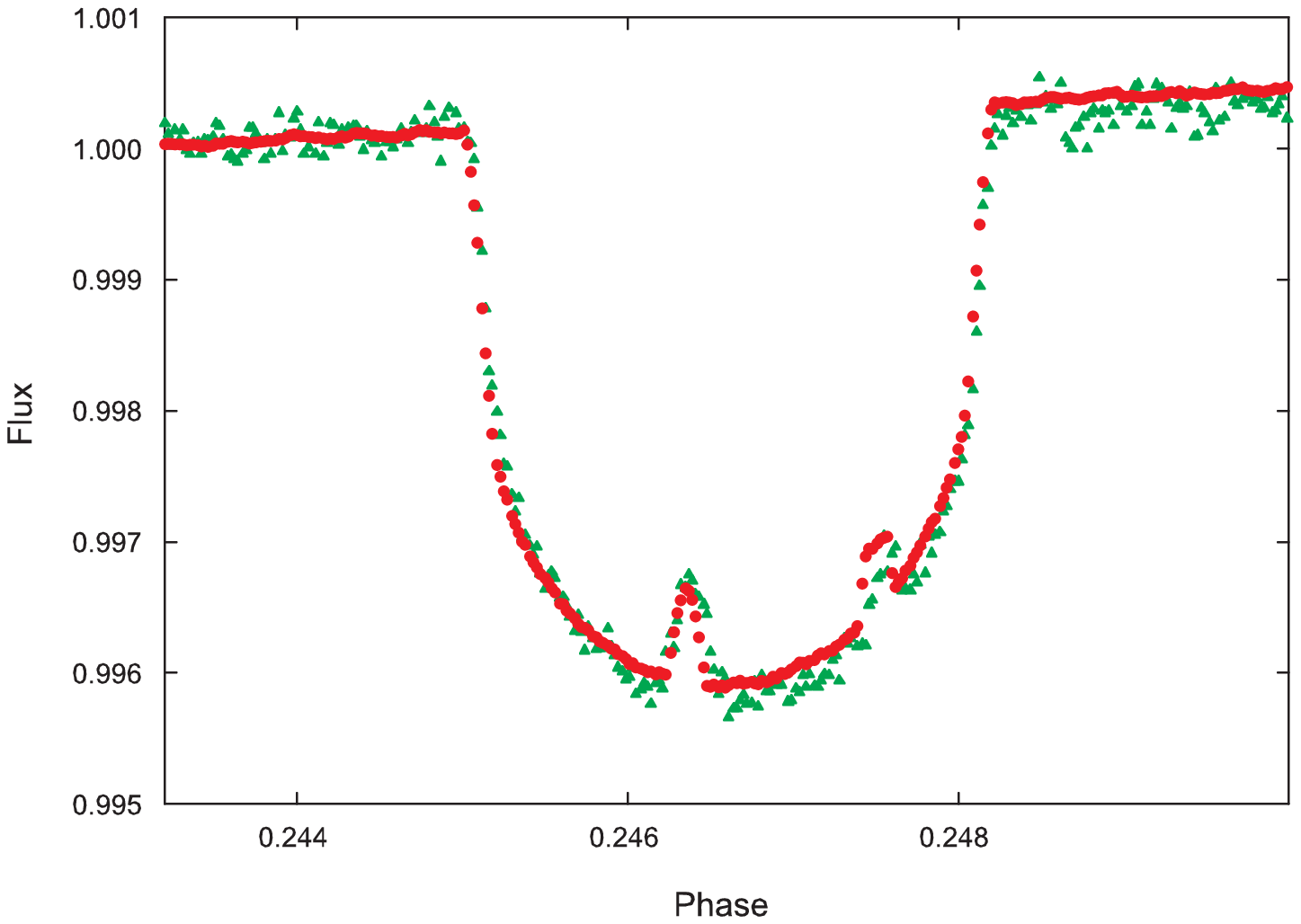}
    }
  }

  \centerline{\hbox{ \hspace{0.0in}
    \epsfxsize=2in
    \epsffile{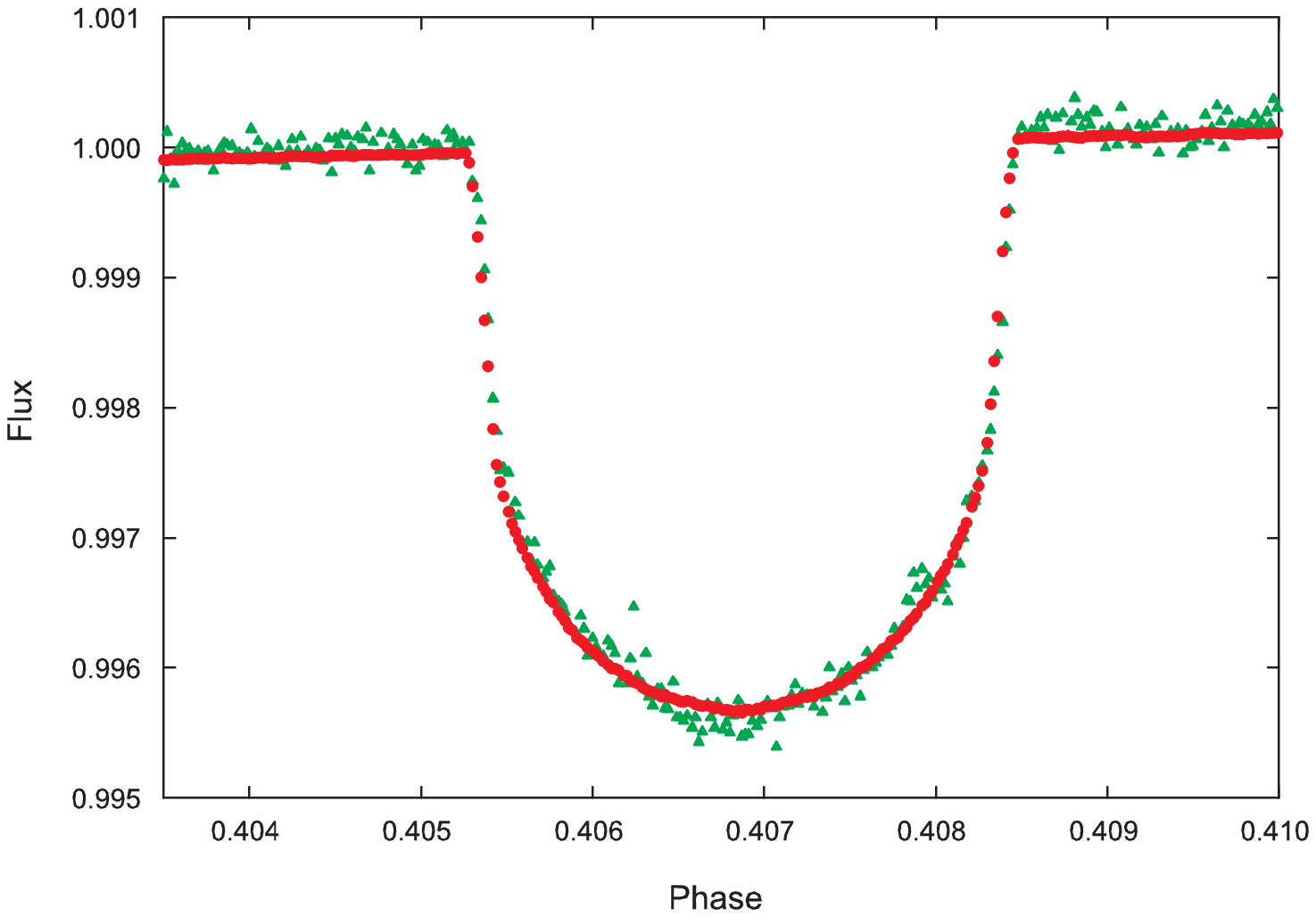}
    \hspace{0.05in}
    \epsfxsize=2in
    \epsffile{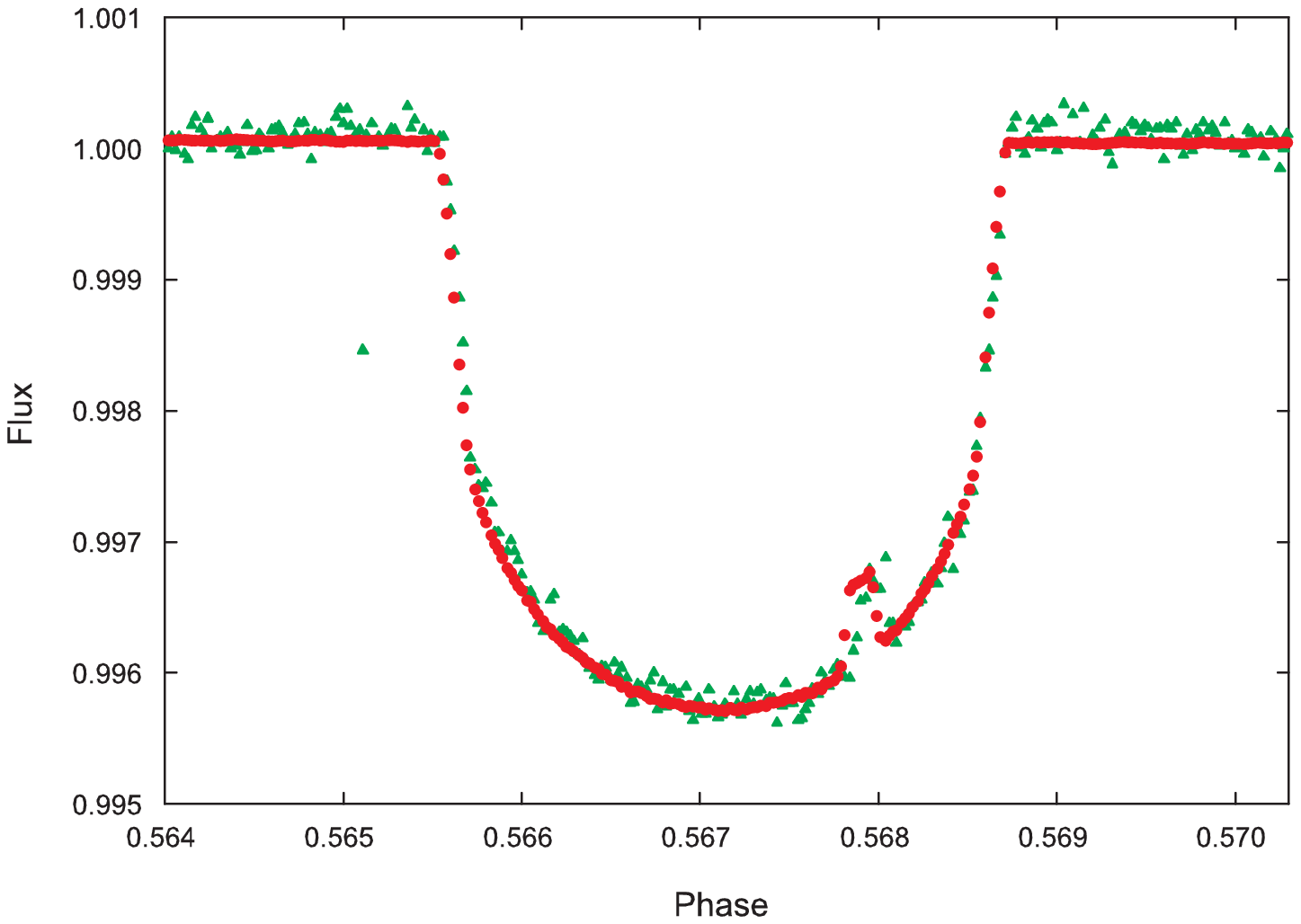}
    }
  }

  \centerline{\hbox{ \hspace{0.0in}
    \epsfxsize=2in
    \epsffile{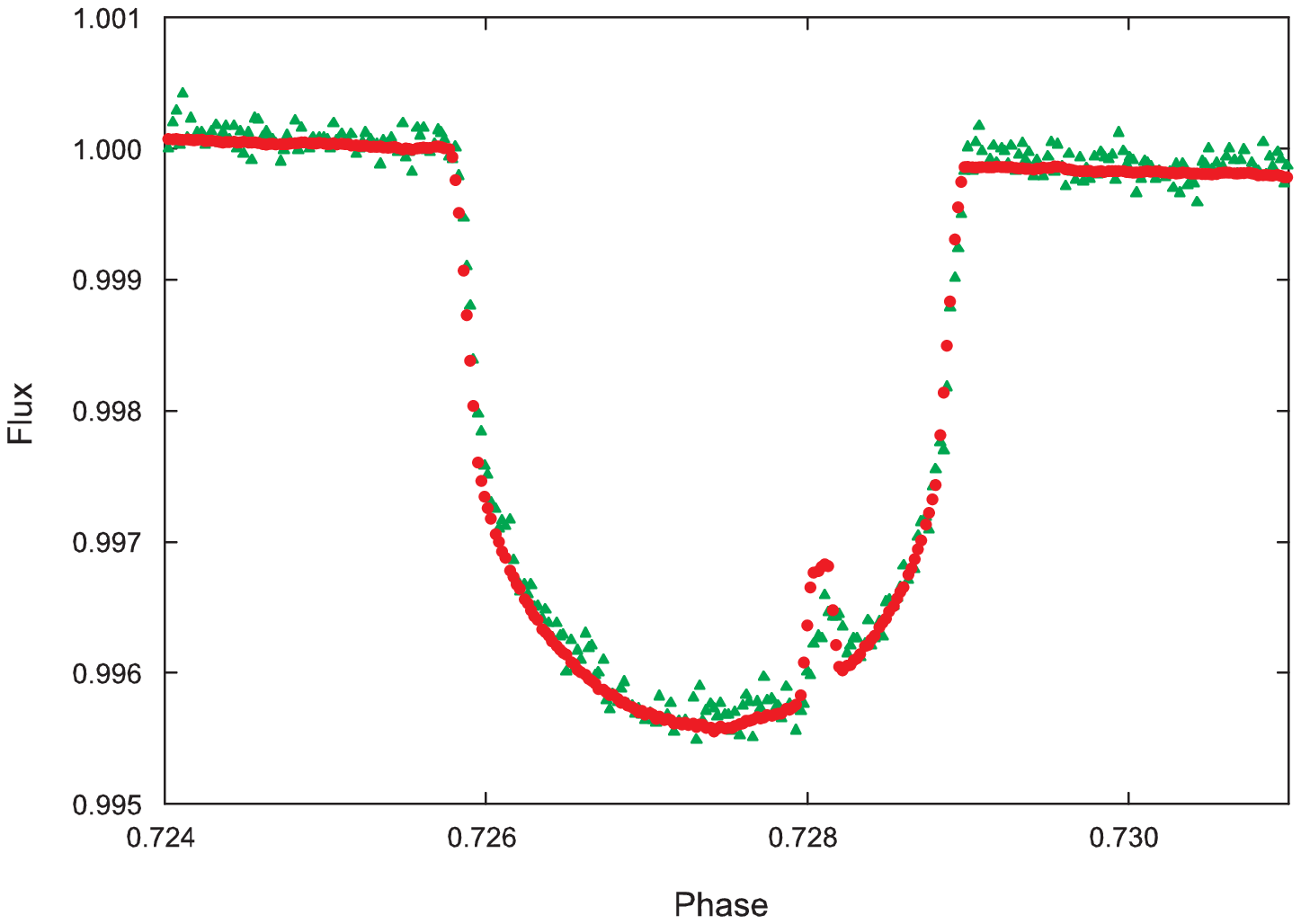}
    \hspace{0.05in}
    \epsfxsize=2in
    \epsffile{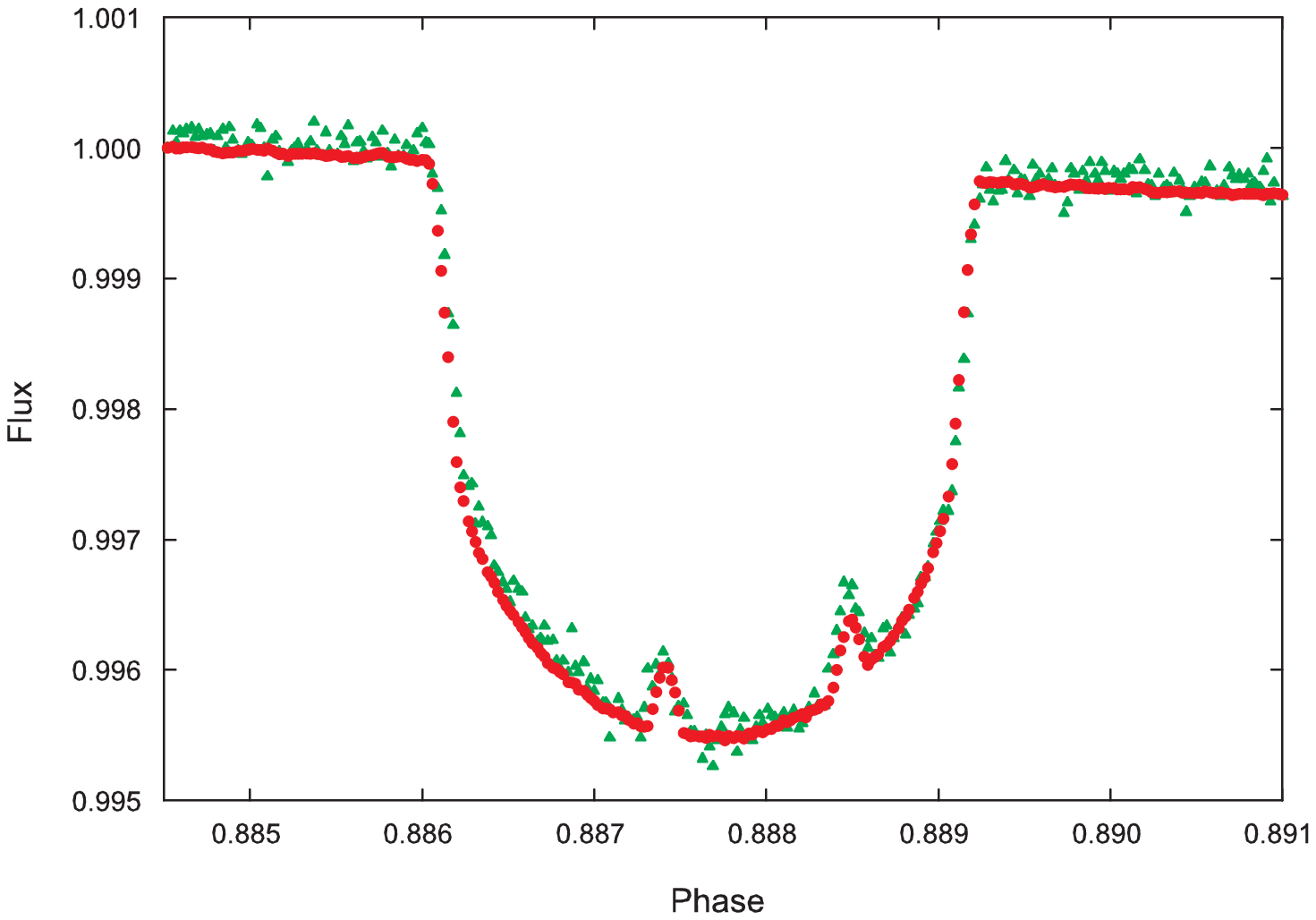}
    }
  }

  \caption{Zoom on the six transits of HAT-P-11 in the figure 5. The flux values are normalized
to one for both observation and SOAP-T. Red dot represent the SOAP-T photometric result and green
triangles are observed flux of HAT-P-11.}
  \label{sub-fig-test}

\end{figure*}

\subsection{HAT-P-11 pole-on solution}

In this section, we examine the pole-on model for the system proposed by
\citet{Sanchis-Ojeda-11b}. We note that here we explore which configuration of spot on the surface of the star would be able to generate the out of transits variation and we did not attempt to reproduce the inside transit features. We generated the light curve of HAT-P-11
using SOAP-T photometry and the parameters of the pole-on model given by these authors.
(see the details of parameters in Table 3). Since the pole-on solution presented by
\citet{Sanchis-Ojeda-11b} is only nearly pole-on
(i.e., the angle of misalignment is $\thicksim 168^\circ$), some spots located close to the equator
of the star can produce large variations outside the transit by disappearing from the
view for a small part of the rotation period. We investigated the effect of different possible sizes
of the spots that were located in different latitudes. Our analysis showed that
the optimized result, whose outside-transit profile is similar to that from the observation,
has to have a spot on $20^\circ$ or $60^\circ$ latitude with a
size equal to 0.2$R_{\ast}$ (see Figure 7). Since the brightness of spot is fixed to zero, it is the minimal size of the spot. For a star with a similar spectral type as HAT-P-11
(i.e., K), this is a very large spot. Although spots with very large sizes have been detected
on K stars \citep{Strassmeier-99}, this situation is unlikely for HAT-P-11.

\begin{figure*}[hbtp]
  \vspace{9pt}

  \centerline{\hbox{ \hspace{0.0in}
    \epsfxsize=2in
    \epsffile{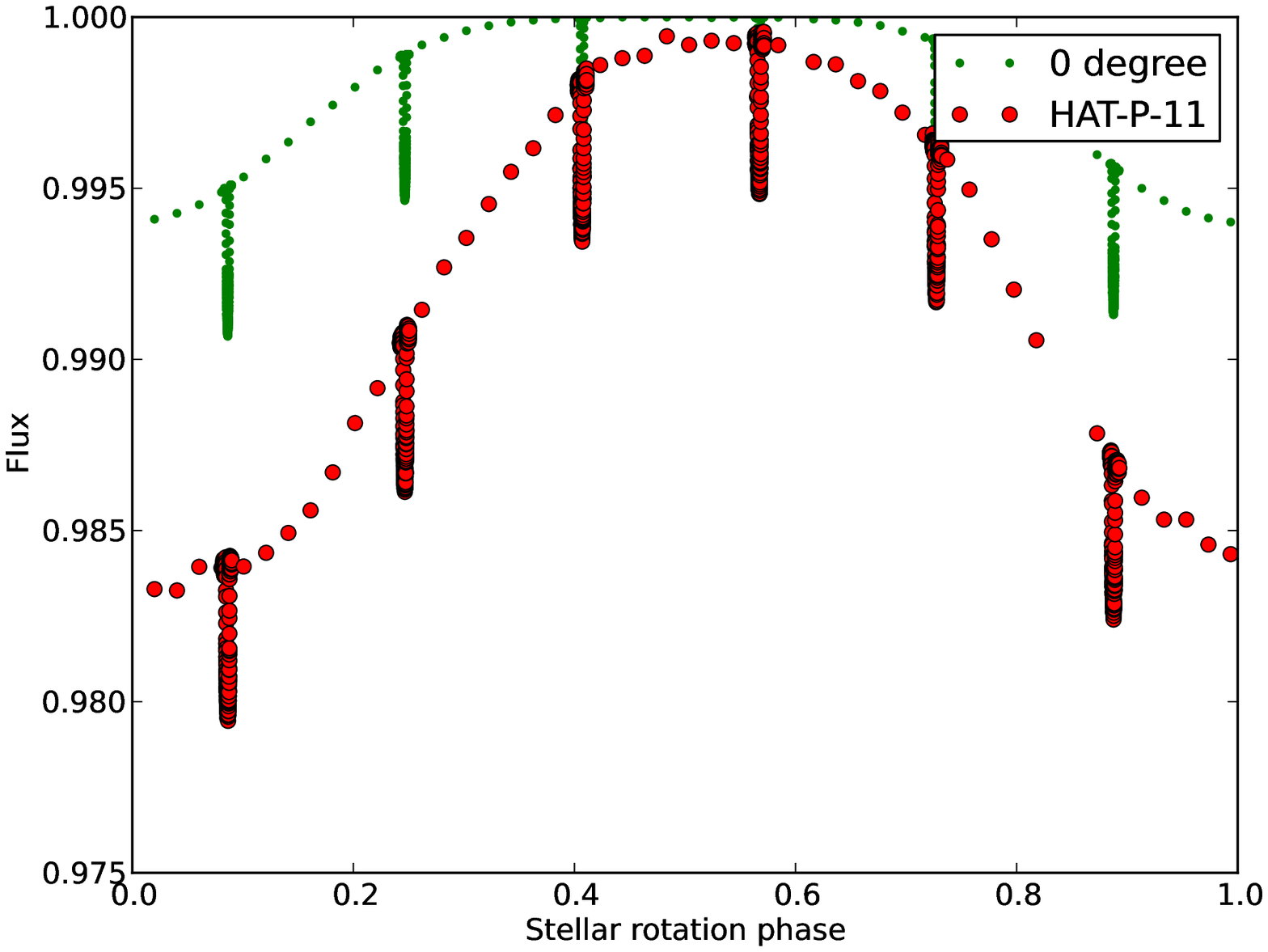}
    \hspace{0.05in}
    \epsfxsize=2in
    \epsffile{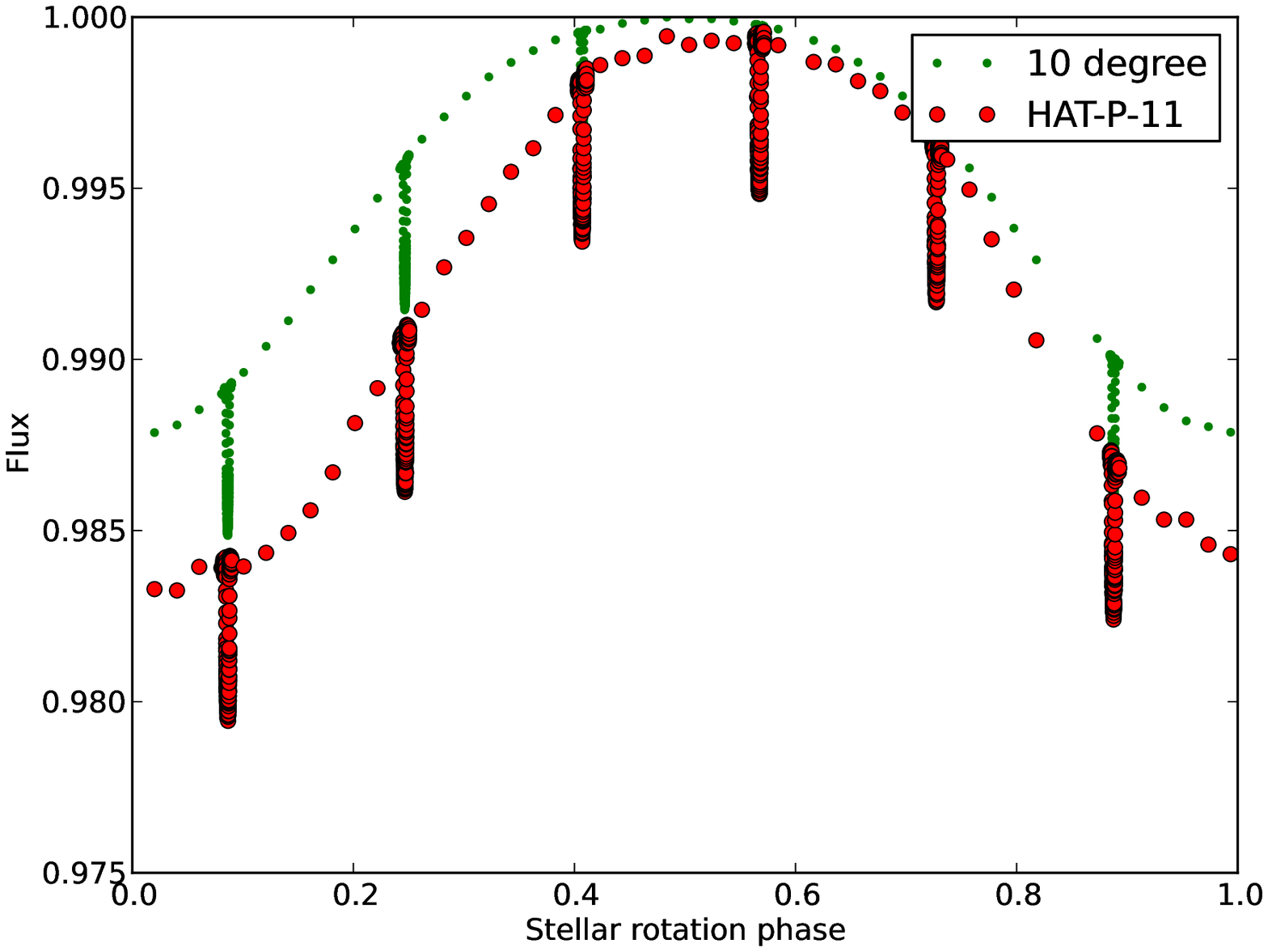}
    \epsfxsize=2in
    \epsffile{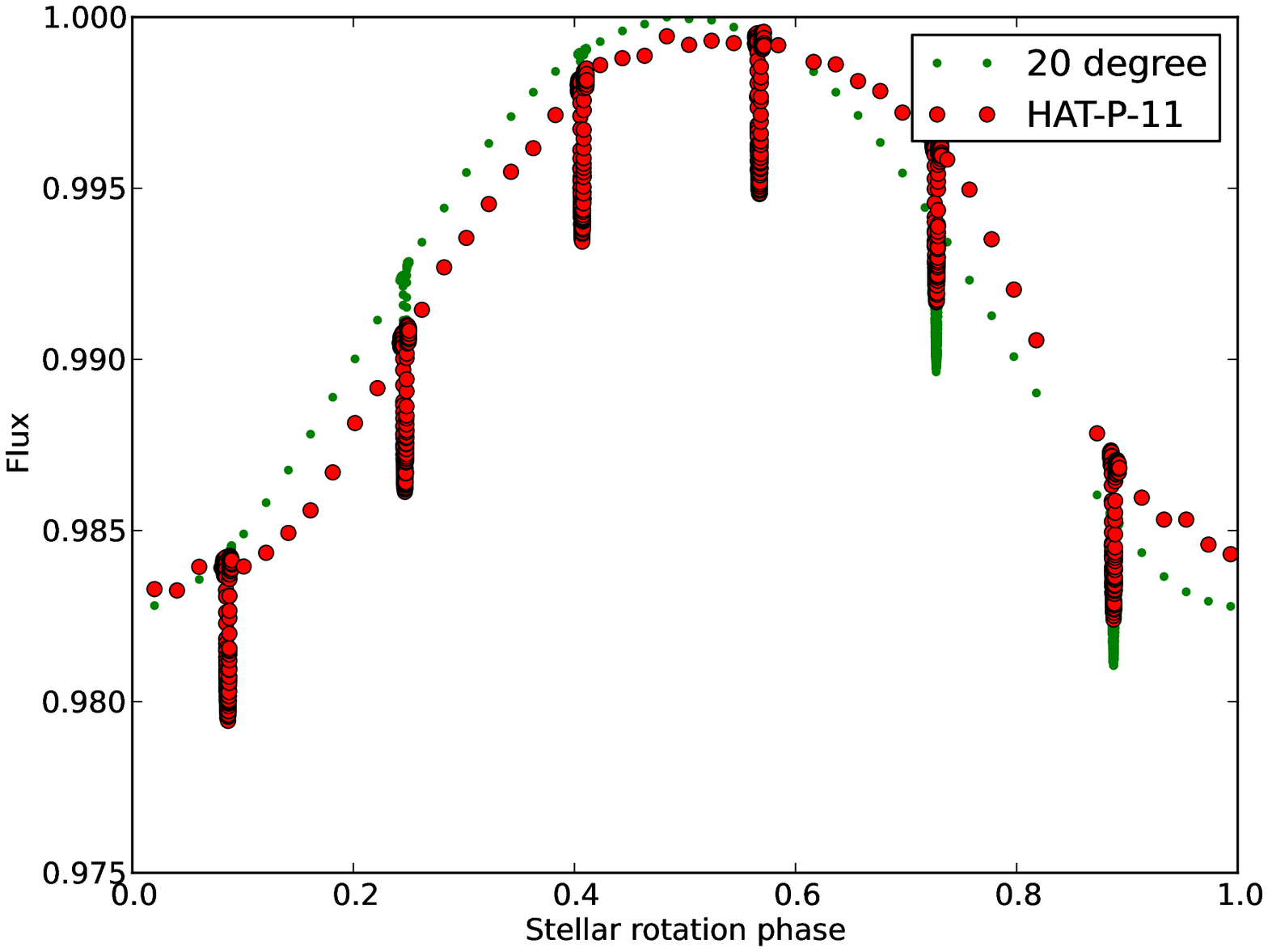}
    }
  }

  \centerline{\hbox{ \hspace{0.0in}
    \epsfxsize=2in
    \epsffile{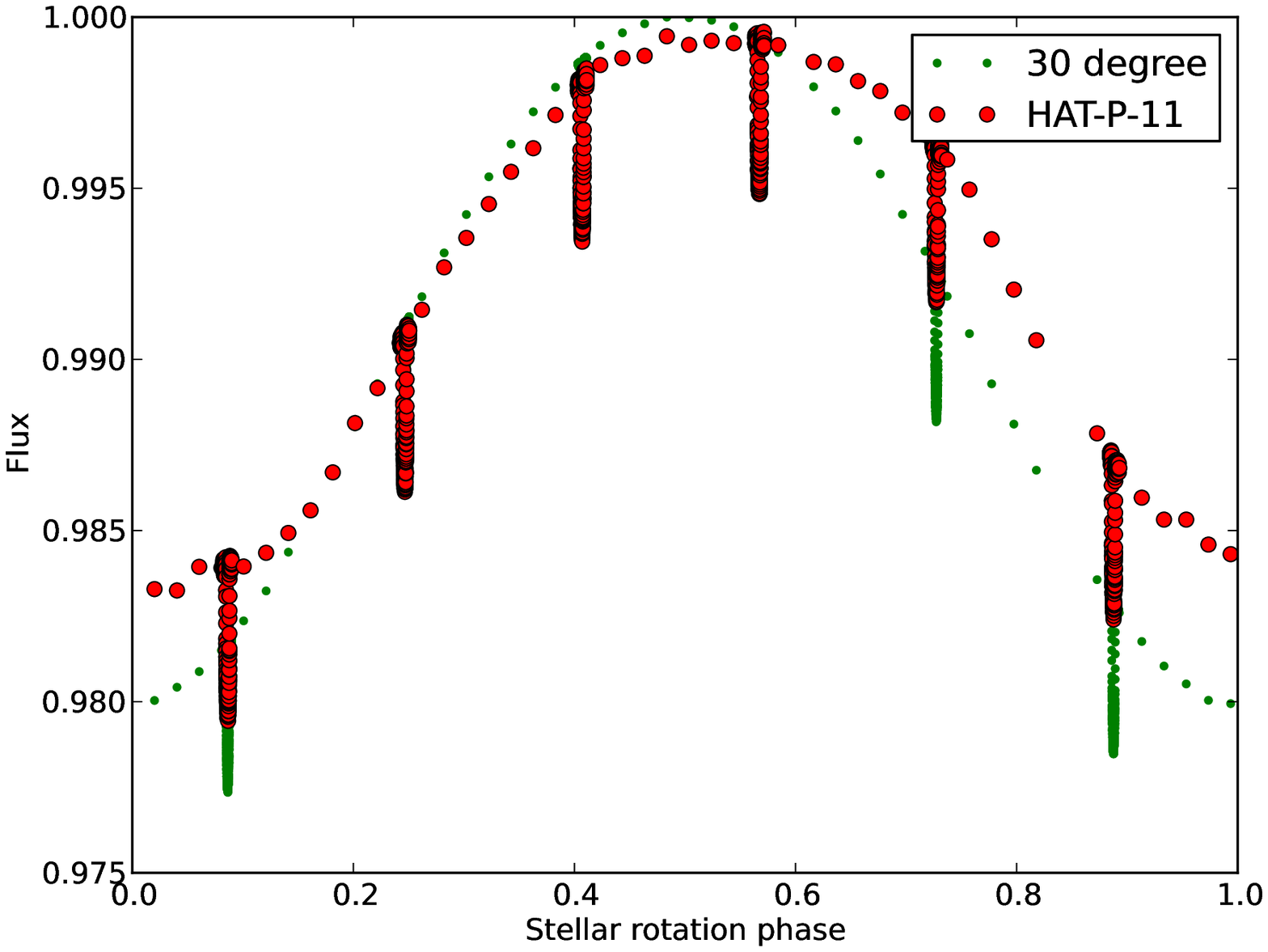}
    \hspace{0.05in}
    \epsfxsize=2in
    \epsffile{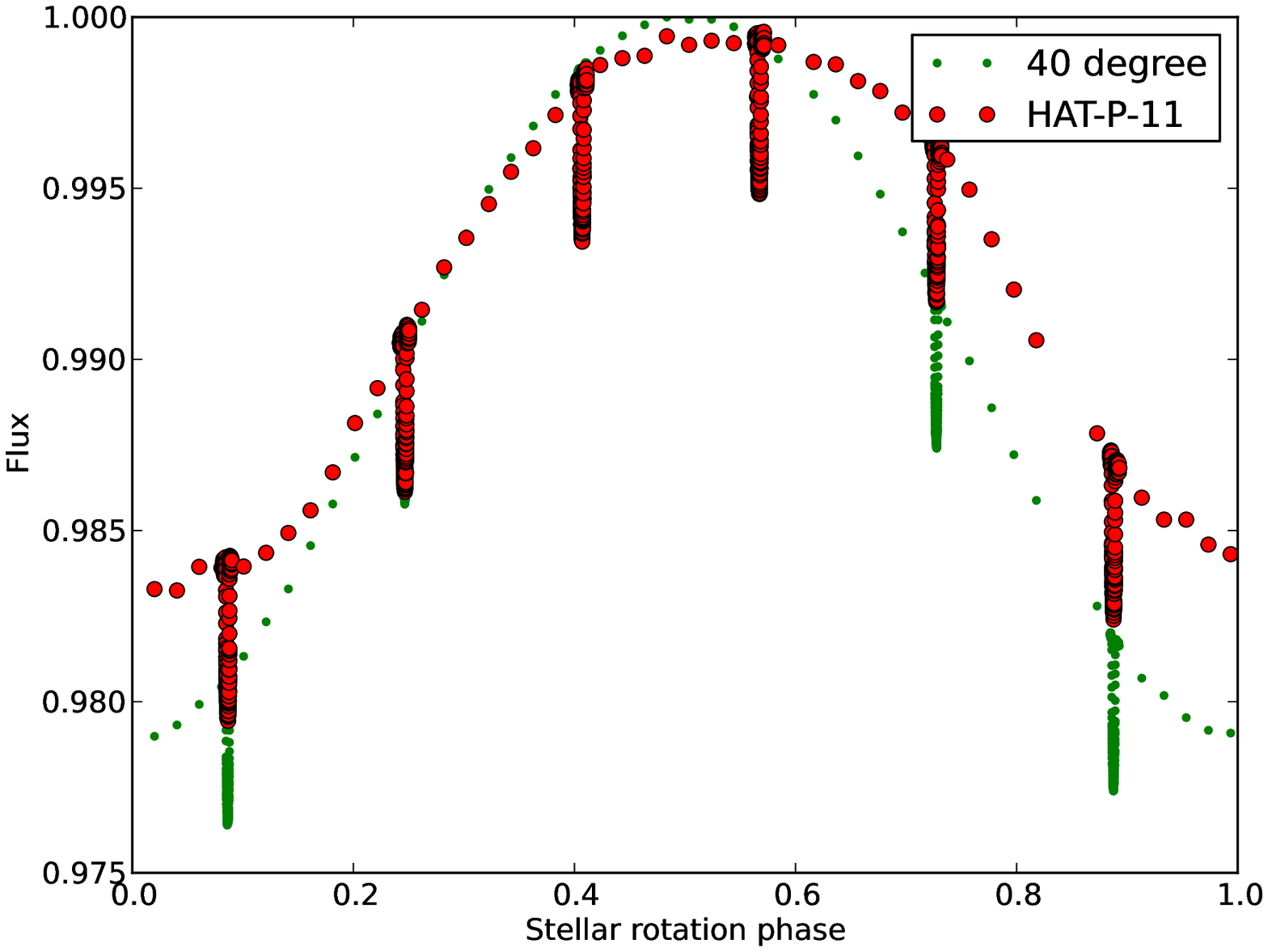}
    \epsfxsize=2in
    \epsffile{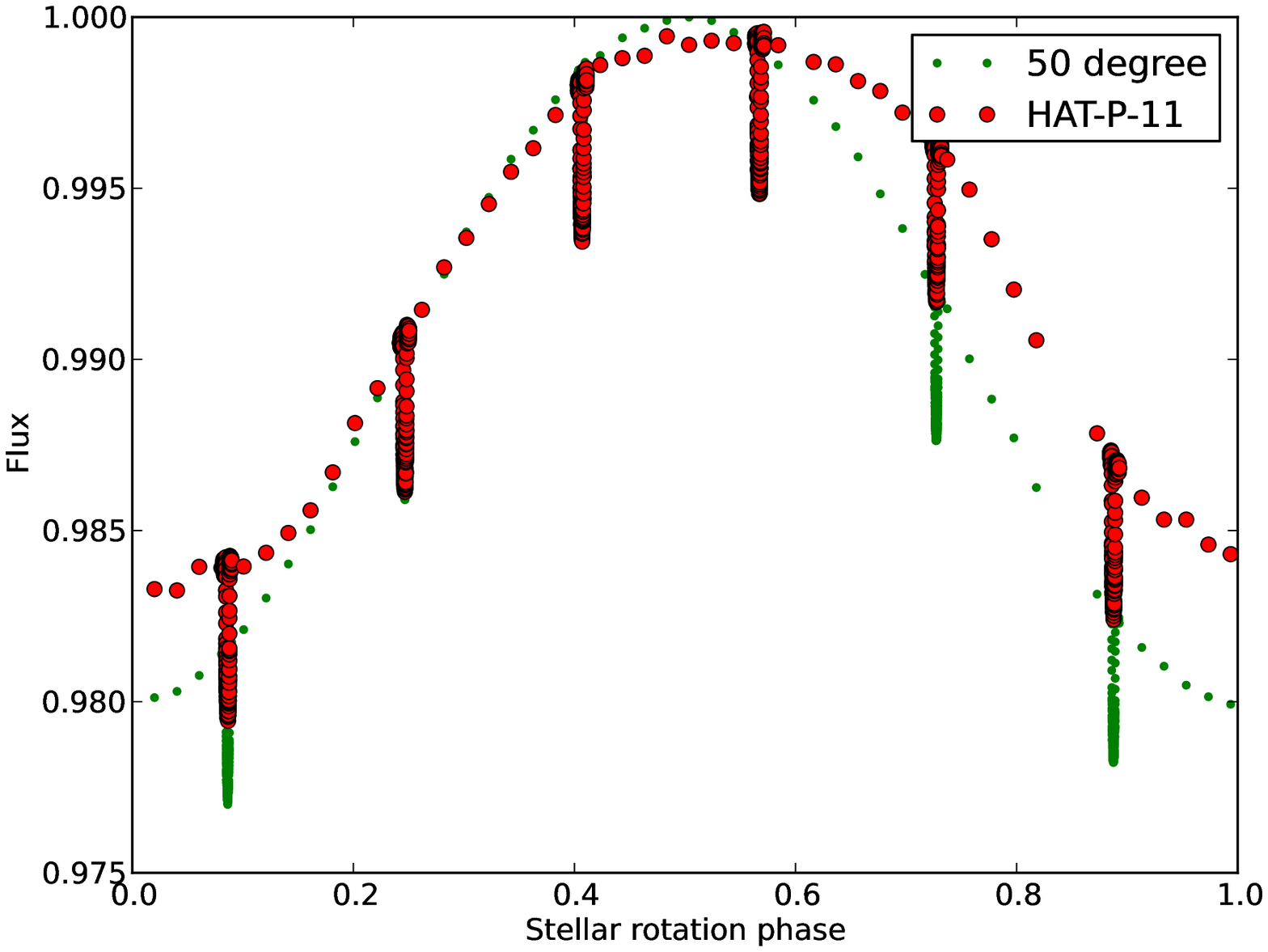}
    }
  }

  \centerline{\hbox{ \hspace{0.0in}
    \epsfxsize=2in
    \epsffile{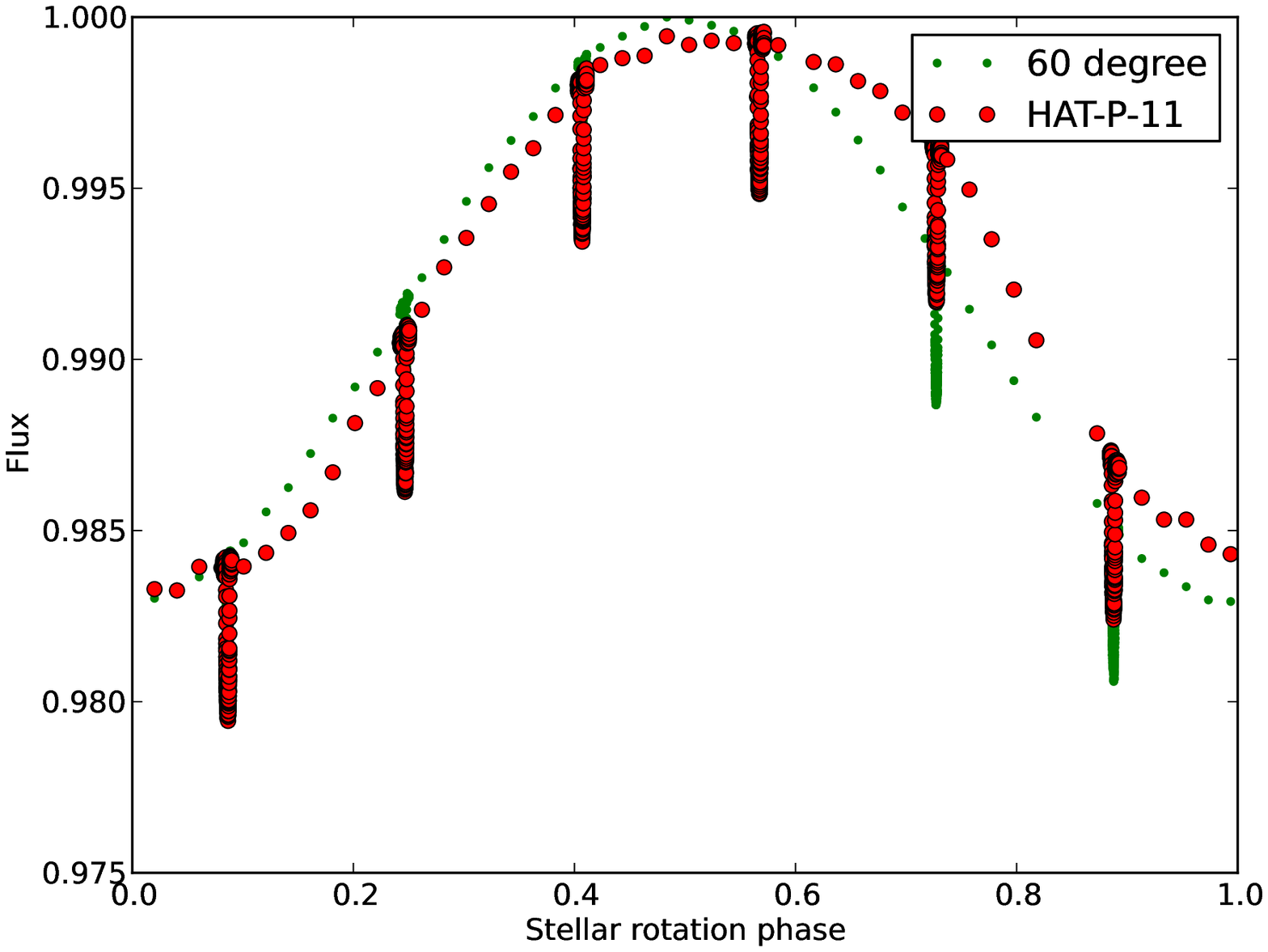}
    \hspace{0.05in}
    \epsfxsize=2in
    \epsffile{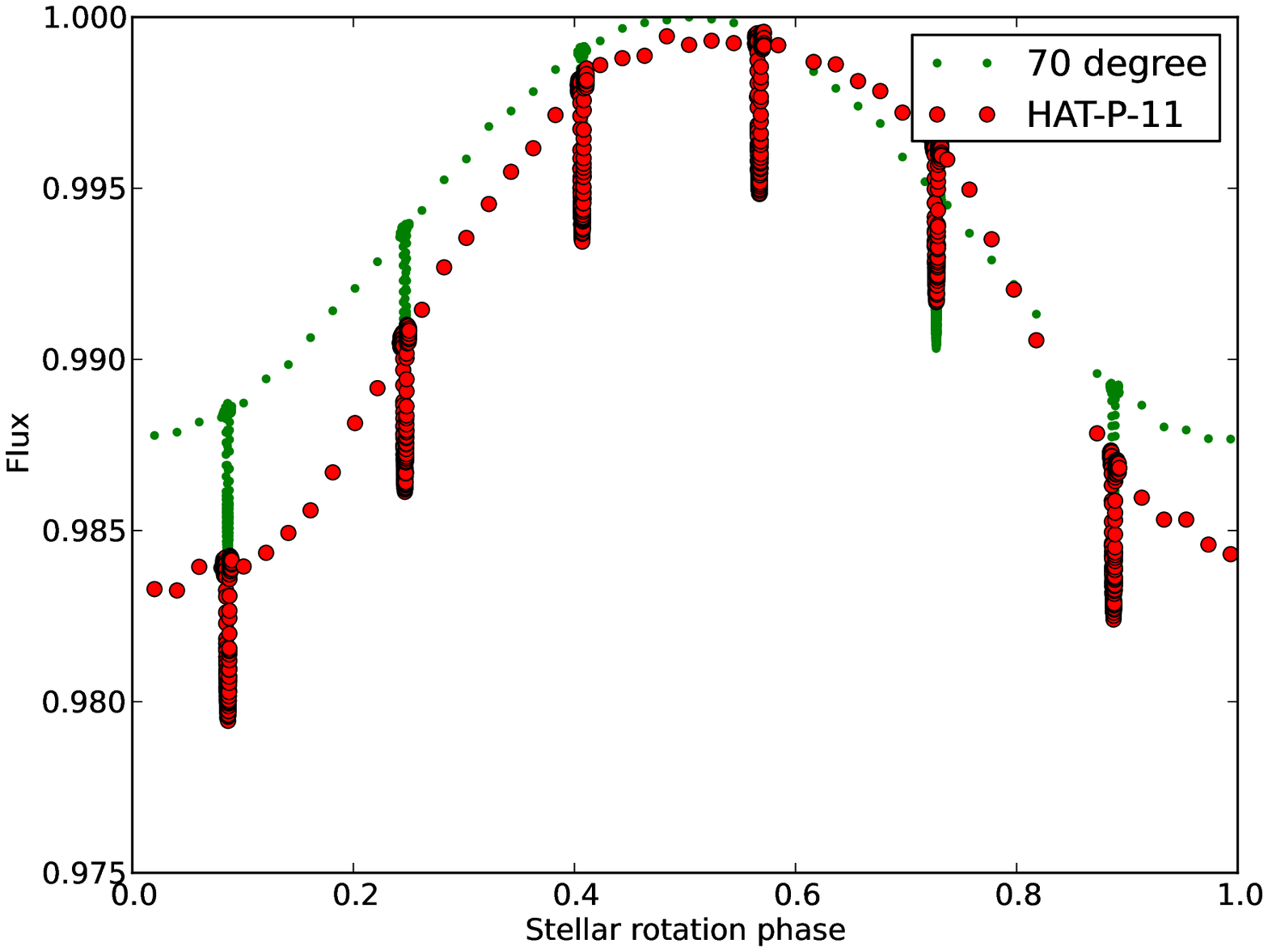}
    \epsfxsize=2in
    \epsffile{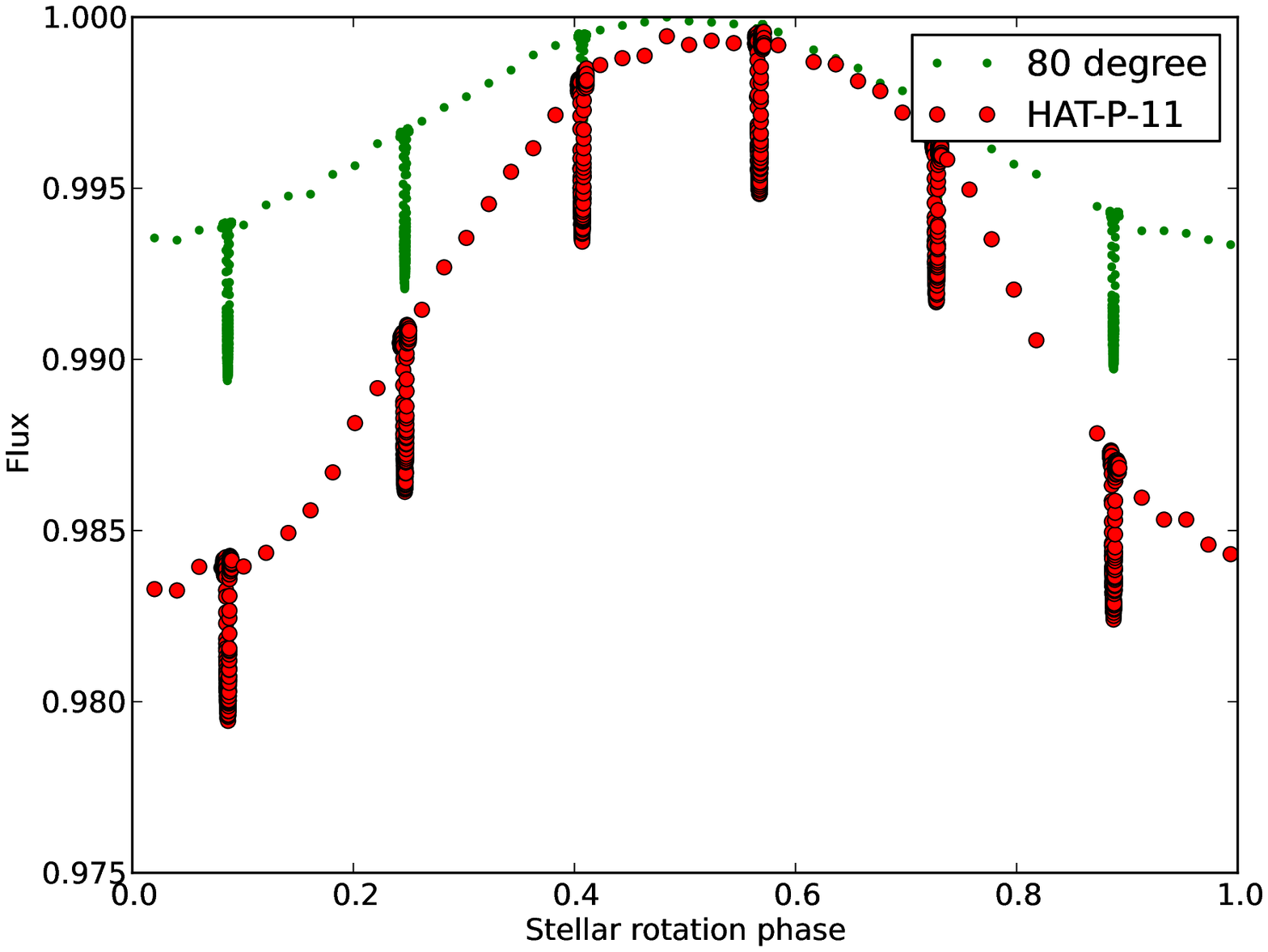}
    }
  }

    \centerline{\hbox{ \hspace{0.0in}
    \epsfxsize=2in
    \epsffile{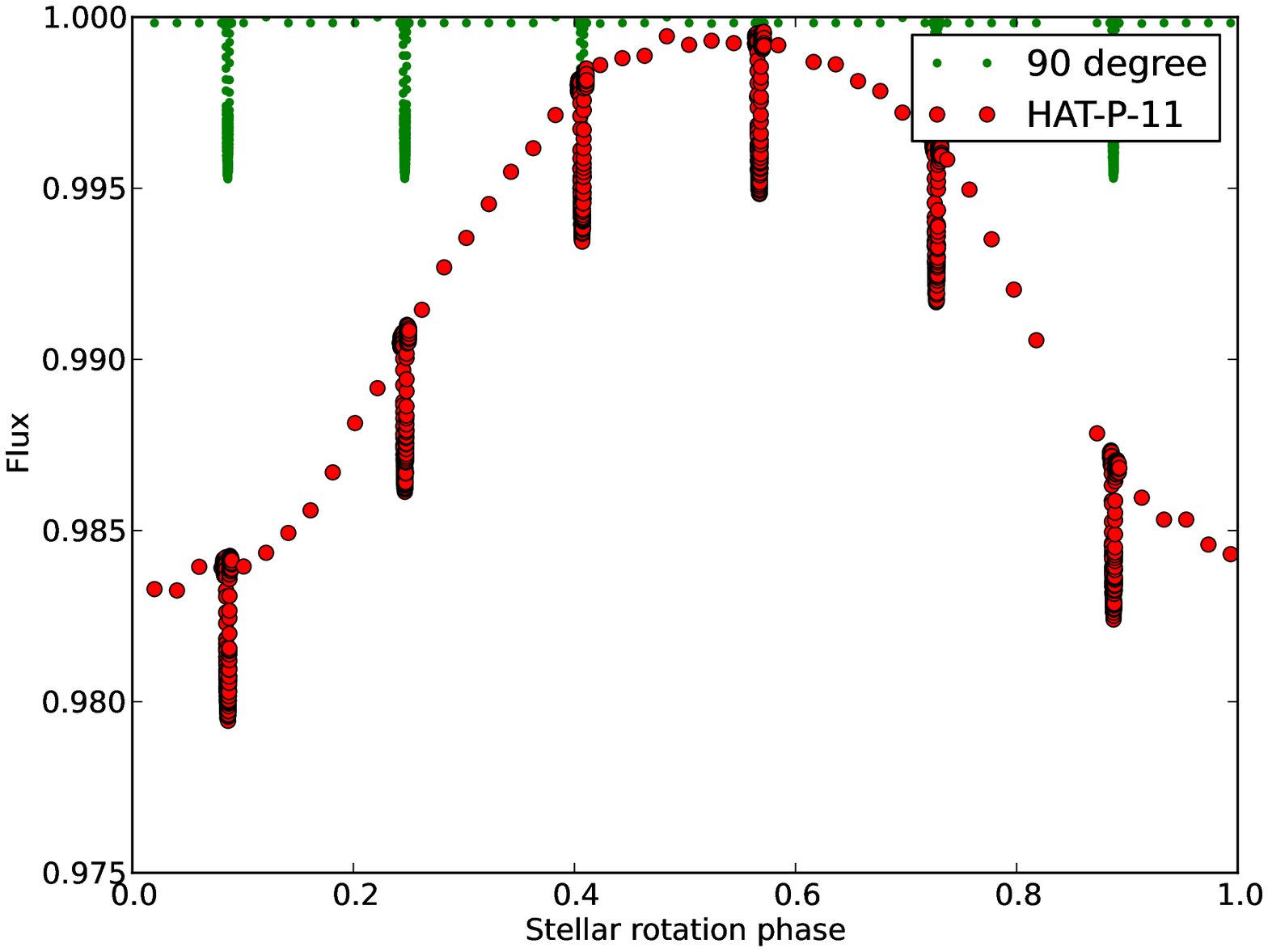}
    }
  }

  \caption{Comparing SOAP-T's photometric result with the observation of HAT-P-11
for the pole-on model of \citet{Sanchis-Ojeda-11b}. Each panel shows different values for the latitude of one spot with
the size of 0.2 radius of star. Green dot is the SOAP-T photometric result and red circle is the observation
of HAT-P-11.}
  \label{sub-fig-test}

\end{figure*}

\section {Conclusion and Perspective}
In this paper, we introduced a new software package that can be used to produce the photometric and RV
signals of a system consisting of a rotating spotted star and a transiting planet where the spots and
planet overlap. We tested the capability of our code
by comparing its results with theoretical models and the results of observations.
We used the HAT-P-11 system and demonstrated that when using the edge-on model of \citet{Sanchis-Ojeda-11b},
SOAP-T is capable of reproducing the same features as obtained from observations
for the outside of transits as well as the anomalies inside the transits.
We also showed that the reconstruction of the large modulation of the outside transit of HAT-P-11
with the pole-on model of \citet{Sanchis-Ojeda-11b} requires an assumption
on the size of spot on the star which may not be realistic.

In using SOAP-T,
the number of free parameters of a system that can be chosen by the user is large. In general,
for a system consisting of a rotating star with one spot and a planet transiting, there are 18 free parameters. As a result, for a system where parameters are
not constrained (e.g., by observation), the
finding of the best solution that fits real observation becomes a complicated task. The
minimization of the reduced chi-squared (the indicator of goodness of the fit) in this case requires
the exploration of a large parameter space. A simple griding of this space turns the calculations
into a time consuming process which may result in the finding of local minima instead of a global
minimum. Efforts are currently underway to investigate different optimization strategies which
can lead to faster and more secure determination of a global minimum.

\begin{acknowledgements}

We acknowledge the support by the European Research Council/European Community under the
FP7 through Starting Grant agreement number 239953,
and by Funda\c{c}\~ao para a Ci\^encia e a Tecnologia (FCT) in
the form of grant reference PTDC/CTE-AST/098528/2008 and SFRH/BPD/81084/2011. NCS also acknowledges the support
from FCT through program Ci\^encia\,2007 funded by FCT/MCTES (Portugal) and
POPH/FSE (EC). NH acknowledges support from the NASA/EXOB program through grant NNX09AN05G
and from the NASA Astrobiology Institute under
Cooperative Agreement NNA09DA77 at the Institute for Astronomy, University
of Hawaii.

Some of the data presented in this paper were obtained from the Multimission Archive at the Space
Telescope Science Institute (MAST). STScI is operated by the Association of Universities for Research in
Astronomy, Inc., under NASA contract NAS5-26555. Support for MAST for non-HST data is provided by the
NASA Office of Space Science via grant NAG5-7584 and by other grants and contracts."

\end{acknowledgements}

\bibliographystyle{aa}
\bibliography{mahlibs}

\end{document}